\documentstyle[prl,aps]{revtex}
\draft
\begin{document}
\begin{titlepage} 
\title{TeV Emission by Ultra-High Energy Cosmic Rays
in Nearby, Dormant AGNs}
\author{Amir Levinson}
\address{School of Physics and Astronomy, Tel Aviv University,
Tel Aviv 69978, Israel} 
\maketitle  
\begin{abstract}
The curvature radiation produced by particles accelerating near the
event horizon of a spinning supermassive black hole, 
threaded by externally supported magnetic field lines is considered.
It is shown that light nuclei suffer catastrophic curvature losses
that limit the maximum energy they can attain to values well below 
that imposed by the maximum potential difference induced by the black 
hole dynamo, unless the curvature
radius of magnetic field lines largely exceeds the gravitational radius.
It is further shown that the dominant fraction of the rotational energy 
extracted from the black hole is radiated in the TeV band.
Given the observed flux of ultra-high energy cosmic rays, and the 
estimated number of nearby supermassive black holes, it is 
expected that if dormant AGNs are the sources of UHECRs, as 
proposed recently by Boldt \& Ghosh, then they should 
also be detectable at TeV energies by present TeV experiments. 
\end{abstract}
\pacs{Pacs Numbers:  98.70.Rz, 98.70.Sa, 95.30.Gv, 98.62.Js}
\end{titlepage}

Energy losses resulting from interactions with the cosmic microwave background 
limit the distance a cosmic ray (CR) of energy $>10^{20}$ eV can traverse 
to less than 50 Mpc \cite{AC94}.  As a consequence, if the origin of the 
ultra-high energy cosmic rays (UHECRs) observed 
is associated with astrophysical objects, rather than decaying supermassive X particles 
as in the top-down scenario \cite{BS99}, then the sources of UHECRs 
must be close by.  Plausible classes of UHECR sources have been 
discussed in the literature, including active galactic nuclei (AGNs) 
\cite{BS87} and gamma-ray bursts \cite{LE93}.

Recently, it has been proposed \cite{BG99} 
that the UHECR observed at Earth may originate from dormant AGNs in the 
local Universe having masses in excess of $10^9 M_{\odot}$. 
The idea being that these objects, although underluminous relative to 
active quasars, contain spinning supermassive black holes that, instead 
of producing luminous radio jets as seen in blazars, serve as 
accelerators of a small number of particles to ultra-high 
energies.  The acceleration mechanism invoked is associated with a BZ 
type process \cite{BZ77};
specifically, individual particles are accelerated by the potential 
difference induced by the rotation of a black hole that is threaded by 
externally produced magnetic field lines, during episodes when breakdown 
of the vacuum does not occur.  In the absence of energy losses, the maximum 
energy a particle can gain by this mechanism is limited to the voltage
drop involved, and is proportional to the product of the black hole 
mass and the strength of magnetic field.   

A recent analysis \cite{M98} indicates that massive dark
objects (MDOs) are present in the centers of nearby galaxies, some of which
have masses in excess of $10^{10} M_{\odot}$.  A plausible interpretation is
that the MDOs are supermassive black holes, and may represent quasar remnants
or dormant AGNs.  This interpretation is further supported by the fact that a 
correlation between the black hole mass and bulge luminosity, similar 
to that found for the sample of galaxies studied in ref. \cite{M98}, has been found
\cite{La98} for a sample of bright quasars, using a completely different 
method.  By applying their model to a list of objects from ref. \cite{M98}, 
of which 14 have compact central masses larger than $10^9 M_{\odot}$, 
Boldt \& Ghosh \cite{BG99} estimate that protons can be 
accelerated up to energies in excess of a few times $10^{20}$ eV.  As 
emphasized by them, in order to account
for the measured flux of UHECRs, an average power of only $\sim 10^{42}$ 
erg s$^{-1}$ is needed to be extracted from a black hole by the 
accelerated cosmic rays, corresponding to a mass loss rate of
order $10^{10}$ g s$^{-1}$, and a loss rate of electric charge that 
constitutes only a small fraction of the total effective 
current required to induce the potential difference across the gap.
 
Acceleration to the maximum energy allowed is possible provided that
radiative losses are sufficiently small. Boldt \& Ghosh \cite{BG99} argued that  
proton energy losses during the acceleration phase due to pair production 
and photomeson production on ambient photons are unimportant by virtue
of the low accretion luminosity anticipated in those objects.  However,
they have not discussed the radiation associated with the acceleration 
process itself.  In the following, we consider the curvature radiation 
produced by the accelerated particles, and show that in the case of light 
nuclei it limits the maximum 
energy attainable to values below the full voltage, unless the average 
radius of curvature of the particle's trajectory exceeds the gravitational 
radius by at least a factor of a few. We further
show that the curvature radiation is emitted predominantly in
the TeV band, and conclude that if dormant AGNs are indeed the sources
of UHECRs, then they should also be detectable by present, ground based experiments 
at TeV energies.

{\it Curvature radiation}: 
The electric potential difference generated by a maximally 
rotating black hole of mass $M=10^9M_9 M_{\odot}$, threaded by 
magnetic field of strength $B=10^4B_4$ Gauss is \cite{TPM}
\begin{equation}
\Delta V\sim 4.4\times10^{20} B_4M_9(h/R_g)^2\ \ \ \ {\rm volts},
\end{equation} 
where $h$ is the gap height, and $R_g=GM/c^2$ is the gravitational radius. 
In the presence of a nonuniform magnetic field, particles accelerated
by this potential difference will suffer energy losses through curvature 
radiation, even if initially they move along magnetic field lines.  Since the 
gyroradius of a proton having energy $\epsilon$, $R_c=\epsilon/eB$, is 
smaller than the gravitational radius: 
\begin{equation}
R_c/R_g= (\epsilon/e\Delta V)e\Delta Vc^2/(eGBM)\simeq (\epsilon/e\Delta V)<1, 
\label{Rc}
\end{equation}
(the gyroradius of an ion of charge $Z$ having energy near the maximum imposed 
by the voltage drop will be larger by a factor of $Z$),
we expect the average radius of curvature of a particle's trajectory
to be of order the curvature radius of magnetic field lines in the gap.
(The curvature radii of different trajectories should span some range though,
 reflecting the different boundary conditions.)  The computations of particles' 
trajectories, even for relatively simple magnetic field topologies, 
are complicated by the fact that the gyroradius at the highest energies
is comparable to the size of the hole, as can be seen from eq. (\ref{Rc}),
and are beyond the scope of this paper.  In what follows, we denote by $\rho$
the average curvature radius of an accelerating ion, and assume that it is 
independent of the ion energy.  The rate of energy 
loss through curvature radiation by a particle of energy 
$\epsilon=mc^2\gamma$ can then be expressed as 
\begin{equation}
P=\frac{2}{3}\frac{e^2c\gamma^4}{\rho^2}.
\label{P}
\end{equation} 
The energy change per unit length of an accelerating ion having charge $Z$ and 
mass $m_i=\mu m_p$ is given by 
\begin{equation}
d\epsilon/ds=eZ\Delta V/h - P/c,
\end{equation}
yielding a maximum acceleration energy,
\begin{equation}
\epsilon_{max}=3\times10^{19}\mu Z^{1/4}M_9^{1/2}B_4^{1/4}
(\rho^2h/R_g^3)^{1/4}\ \ \ {\rm eV},
\label{Emx}
\end{equation}
where eq. (\ref{P}) has been used.  Consequently, only a fraction 
\begin{equation}
\eta=0.1 \mu M_9^{-1/2}(ZB_4)^{-3/4}(\rho/R_g)^{1/2}(h/R_g)^{-7/4};\ \ \ \ \eta\le1,
\label{eta}
\end{equation}
of the potential energy available will be released as UHECRs;
the rest will be radiated in the form of curvature photons.
For the most massive MDOs listed in table 2 of ref. \cite{M98} ($M_9>20$)
we obtain $\epsilon_{max}\sim 1.5\times10^{20}\mu
(ZB_4\rho^2h/R_g^3)^{1/4}$ eV,
and $\eta\sim 0.02\mu (ZB_4)^{-3/4}(\rho/R_g)^{1/2}(h/R_g)^{-7/4}$. 
Thus $B_4\rho^2h/R_g^3\ge 20$ is required in order to accelerate a proton
($Z=\mu=1$) to energies $\ge 3\times10^{20}$ eV in these systems, corresponding 
to $\eta=10^{-3}(\rho/R_g)^2(h/R_g)^{-1}$.
The requirement on heavier nuclei is more relaxed.  Estimates of the maximum 
value of the horizon threading magnetic field strength yield $B_4\le1$ \cite{TPM}, 
and according to recent numerical simulations $B_4$ may be well below unity 
\cite{GA97}.
Consequently, acceleration of light nuclei to the highest energies measured by current
experiments requires $\rho$ to be larger than $R_g$ by a factor of at least a 
few (assuming $h \sim R_g$).  The values of $B_4$ obtained by Boldt \& Ghosh 
\cite{BG99} for the Magorian et al. sample \cite{M98}, assuming equipartition 
between the magnetic field in the vicinity of the horizon and the matter 
infalling into the center, lie in the 
range between 0.1 to 1, given the estimated mass loss rate of the galaxy bulge.   

The spectrum produced by the curvature radiation of a single ion 
will peak at an energy
\begin{equation}
\epsilon_{\gamma max}=1.5\gamma^3\hbar c/\rho=1.6\times10^{-7}\epsilon_{max}\mu^{-1}
(ZB_4)^{1/2}(h/R_g)^{1/2}
=5 M_9^{1/2}(ZB_4)^{3/4}(\rho^2 h^3/R_g^5)^{1/4}\ \ \  {\rm TeV},
\label{Eg}
\end{equation}
and is a power law $I(\epsilon_{\gamma})\propto (\epsilon_{\gamma})^{1/3}$
below the cutoff.  The overall spectrum of curvature photons would depend
on the energy distribution of the accelerating particles, and is expected
to be somewhat softer below the peak.  For $\epsilon_{max}=3\times 10^{20}$
eV and $h\sim R_g$ we obtain from eq. (\ref{Eg}), $\epsilon_{\gamma max}\simeq50
\mu^{-1}(ZB_4)^{1/2}$ TeV.

The number of TeV photons 
per proton produced in the process is roughly
\begin{equation}
n_{\gamma}\sim eZ\Delta V/\epsilon_{\gamma max}=10^8 M_9^{1/2}
(ZB_4)^{1/4}(h/R_g)^{5/4}(\rho/R_g)^{-1/2}.
\end{equation}
The mean free path to pair creation of a photon 
moving at an angle $\chi$ to the magnetic field is $l\simeq
95(B_4\sin\chi)^{-1}e^q$, with $q=1.3\times 10^3 M_9^{-1/2}B_4^{-7/4}
\sin\chi^{-1}(\rho^2h^3/R_g^5)^{-1/4}$ and, therefore, the radiated photons
escape the system freely for the range of parameters considered here.
However, when $q$ becomes sufficiently small (at $q$ of about 50 
one photon per proton will be converted into an electron-positron pair),
a pair cascade may be initiated with high enough probability to lead to
a breakdown of the gap.

{\it Observational consequences}:
As shown recently \cite{Wa95b}, the observed CR spectrum above $10^{19}$ eV, as 
measured by the AGASA \cite{Bi94} and Fly's Eye \cite{Ha94} experiments, can be 
accounted for by 
a homogeneous cosmological distribution of CR sources with power law spectra and 
energy production rate of $1.5\times 10^{37}$ ergs Mpc$^{-3}$ s$^{-1}$. 
Given this energy production rate, and
denoting by $n_{CR}$ the density of objects contributing to the observed
CR flux in this energy range, the average power released in the form of UHECR 
by a single source can be expressed as 
\begin{equation}
L_{CR}=2\times 10^{42}\left(\frac{n_{CR}}{10^{-4}Mpc^{-3}}\right)^{-1}\ \ 
{\rm erg\  s^{-1}}.
\label{LcR}
\end{equation} 
Employing eq. (\ref{LcR}), one finds that the average TeV flux emitted by
a single CR source at a distance of $D=50 D_{50}$ Mpc is 
\begin{equation}
F_{\gamma}\simeq 10^{-12}\left(\frac{n_{CR}}{10^{-4}Mpc^{-3}}
\right)^{-1}\eta^{-1}D_{50}^{-2}\ \ \ \ {\rm erg\ cm^{-2}\ s^{-1}}, 
\label{Fg}
\end{equation}
where $\eta<1$ is the UHECR production efficiency defined 
in eq. (\ref{eta}).  As seen from eq. (\ref{eta}), $\eta$ should be of order unity
if the particles accelerated by the black hole dynamo are predominantly 
heavy nuclei.  There is, however, evidence that the CR composition changes from 
heavy nuclei below the ankle (at energy of $\sim 5\times 10^{18}$ eV) to light 
nuclei above it \cite{Bi94}.  Assuming a protonic CR composition above the ankle,
we obtain values of $\eta$ between 10$^{-1}$ and 10$^{-3}$ for 
the range of parameters considered above.     
 
The density of black holes above a certain mass can be estimated using the 
correlation between bulge luminosity and MDO mass found by Magorian 
et al. \cite{M98}, and an appropriate luminosity function of 
nearby galaxies to correct for the incompleteness of their sample.
However, as seen from eq. (\ref{Emx}), the maximum energy a particle can attain 
depends on a combination of parameters and not solely on the mass. Therefore, 
the black hole mass above which particles can accelerate to the required energies
is uncertain, and since the MDOs having relevant masses lie on the bright end of 
the luminosity function, this uncertainty in mass translates to a large 
uncertainty in $n_{CR}$.  For a reasonable choice of parameters we anticipate that 
$M_9$ of at least a few would be required in order to account for UHECR energies
observed.  Using the luminosity function measured by  Efstathiou et al. 
\cite{ERB88}, we estimate that 
the density of MDOs having masses $M_9>1$ is of order a few times 
$10^{-4}$ Mpc$^{-3}$.  An estimate based on a k-band luminosity function 
measured more recently \cite{Lo00} yielded a similar value.

The threshold flux for a 5$\sigma$ detection of gamma-rays above TeV by current 
TeV experiments is $\sim 5\times10^{-12} t_{day}^{-1/2}$ erg s$^{-1}$ cm $^{-2}$,
where $t_{day}$ is the exposure time measured in days \cite{We96}.  With the above 
estimation of $\eta$ and $n_{CR}$, we expect that at least some fraction 
of the UHECR sources will be detectable by present TeV experiments, provided
that the TeV photons escape the system.  Conceivable sources of opacity that 
may give rise to attenuation of the TeV flux are considered next.
    
The curvature photons produced near the black hole can be absorbed through 
pair production on IR photons in the galaxy.    
The corresponding optical depth at a given energy depends on the spectrum 
of IR photons and the energy dependence of the cross section.
To an order of magnitude it is given by, 
$\tau_{\gamma\gamma}\simeq \sigma_{PP} n_{IR} R$, where $R$ is the size
of IR emission region - of order several kpc as inferred from the light 
profiles, $n_{IR}$ is the number density of IR photons, and  
$\sigma_{PP}$ is the pair production cross section, given approximately 
by $\sigma_{PP}\simeq 0.2 \sigma_T$ at energies near the threshold.  Using 
this approximation one finds that TeV photons would escape the galaxy provided 
that the IR luminosity 
at energy $\epsilon_{IR}=(m_ec^2)^2/\epsilon_{\gamma}$ satisfies:
\begin{equation}
L_{IR}(\epsilon_{\gamma}^{-1})<3\times 10^{45} \left(\frac{R}{{\rm 3\ kpc}}\right)
\left(\frac{\epsilon_{\gamma}}{{\rm 1\ TeV}}\right)^{-1}\ \ \ {\rm erg\ s^{-1}}.
\label{Lir}
\end{equation}
The integrated IR luminosity implied by eq. (\ref{Lir}) would presumably be larger.
Condition (\ref{Lir}) is satisfied in most cases.
A central continuum source, if present, would also contribute a pair production 
opacity.   The IR emission may arise from a cold accretion disk 
or dust reprocession, and is likely to originate from small
radii, of order 10 to 100 $R_g$ in these low luminosity objects. 
Adopting a size of 10 times the gravitational radius of 
a $10^{10}M_{\odot}$ black hole for the IR emission region, 
we find that the TeV flux will be strongly attenuated if the 
corresponding IR luminosity exceeds
$\sim 10^{40}(\epsilon_{\gamma}/{\rm 1\ TeV})^{-1}$ ergs s$^{-1}$.  This is 
slightly below the luminosity inferred for low-luminosity AGNs \cite{Ho99},
and comparable to upper limits on the luminosity of the point source in 
elliptical galaxies \cite{FR95}.

In conclusion, it has been shown that particles accelerating near the horizon of 
a spinning suppermassive black hole threaded by externally supported 
magnetic field lines, suffer 
severe energy losses through curvature emission.  The curvature losses limit
the maximum energy attainable by light nuclei to values well below that 
imposed by the voltage drop.  The major fraction of the energy extracted from the 
rotating hole is radiated in the TeV band, with a rather hard spectrum that extends
well beyond 10 TeV.  Given the energy flux of cosmic rays 
above $10^{19}$ eV, as reported recently by current CR experiments, and an 
estimate of the density of supermassive black holes in the universe, it is 
concluded that if dormant AGNs are the sources of the ultra-high energy 
cosmic rays, then they should be detectable by current TeV experiments.

I thank Ari Laor, Dan Maoz, Eli Waxman, David Eichler, Amiel Sternberg, 
and Hagai Netzer for discussions and useful comments.  Support from the 
Israel Science Foundation is acknowledged.

%\newpage

\end{document}